# Three-point correlation function of galaxy clusters in cosmological models: a strong dependence on triangle shapes

Y. P. Jing,[1] G. Börner[1] and R. Valdarnini[2]

[1] *Max-Planck-Institut für Astrophysik, Karl-Schwarzschild-Strasse 1, 85748 Garching, Germany*
*E-mail: jing@mpa-garching.mpg.de; grb@mpa-garching.mpg.de*
[2] *SISSA-International School for Advanced Studies, Via Beirut 2, 34014 Trieste, Italy. E-mail: valda@tsmi19.sissa.it*

**ABSTRACT**
In this paper, we use large P$^3$M N-body simulations to study the three-point correlation function $\zeta$ of clusters in two theoretical models. The first model (LCDM) is a low-density flat model of $\Omega_0 = 0.3$, $\Lambda_0 = 0.7$ and $h = 0.75$, and the second model (PIM) is an Einstein-de-Sitter model with its linear power spectrum obtained from observations. We found that the scaled function $Q(r, u, v)$, which is defined as the ratio of $\zeta(r, ru, ru+rv)$ to the hierarchical sum $\xi(r)\xi(ru)+\xi(ru)\xi(ru+rv)+\xi(ru+rv)\xi(r)$ (where $\xi$ is the two-point correlation function of clusters), depends weakly on $r$ and $u$, but very strongly on $v$. $Q(r, u, v)$ is about 0.2 at $v = 0.1$ and 1.8 at $v = 0.9$. A model of $Q(r, u, v) = \Theta 10^{1.3v^2}$ can fit the data of $\zeta$ very nicely with $\Theta \approx 0.14$. This model is found to be universal for the LCDM clusters and for the PIM clusters. Furthermore, $Q(r, u, v)$ is found to be insensitive to the cluster richness. We have compared our N-body results with simple analytical theories of cluster formation, like the peak theories or the local maxima theories. We found that these theories do not provide an adequate description for the three-point function of clusters. We have also examined the observational data of $\zeta$ presently available, and have not found any contradiction between the observations and our model predictions. The $v$-dependence of $q$ in a projected catalogue of clusters is shown to be much weaker than the $v$-dependence of $Q$ found in the three-dimensional case. This is probably the reason why the $v$-dependence of $Q$ has not been found in an angular correlation function analysis of the Abell catalogue. The $v$-dependence found in this paper might be an important feature of clusters formed in the Gaussian gravitational instability theories. Therefore it would be important to search for the $v$-dependence of $Q$ in redshift samples of rich clusters.

## 1 INTRODUCTION

The observational fact that the two-point correlation function of rich clusters of galaxies has an amplitude much higher than that of galaxies (Bahcall & Soneira 1983; Klypin & Kopylov 1983), has been explained by Kaiser (1984) as a consequence of the hypothesis that rich clusters form at high peaks of the underlying Gaussian fluctuation field on the cluster mass scale. In his paper, Kaiser derived a relation between the two-point correlation function $\xi_{pk}$ of peaks and the two-point function $\xi_m$ of the underlying mass. This derivation has been refined and generalized to the three-point correlation function $\zeta_{pk}$ of peaks by many authors (e.g. Politzer & Wise 1984; Matarrese, Lucchin & Bonometto 1986; Jensen & Szalay 1986). For a Gaussian perturbation field and in the limits of high sharp peaks and weak underlying clustering ($\xi_m \ll 1$), the three-point function $\zeta_{pk}$ is found to be

$$\zeta_{pk}(r_{12}, r_{23}, r_{31}) = Q_h \left[ \xi_{pk}(r_{12})\xi_{pk}(r_{23}) + \xi_{pk}(r_{23})\xi_{pk}(r_{31}) + \xi_{pk}(r_{31})\xi_{pk}(r_{12}) \right] + Q_k \xi_{pk}(r_{12})\xi_{pk}(r_{23})\xi_{pk}(r_{31}) \qquad (1)$$

with $Q_h = Q_k = 1$. For this case, $\zeta_{pk}$ is said to have the Kirkwood form. If $Q_k$ is zero, $\zeta_{pk}$ is said to have the Hierachical form. Szalay (1988) relaxed the limit of high sharp peaks and examined $\zeta_{pk}$ for a general threshold function. Under the limit of weak underlying clustering, he showed that the cubic term $Q_k \xi_{pk}(r_{12})\xi_{pk}(r_{23})\xi_{pk}(r_{31})$ always accompanies the hierarchical term $Q_h[\xi_{pk}(r_{12})\xi_{pk}(r_{23}) + \xi_{pk}(r_{23})\xi_{pk}(r_{31}) + \xi_{pk}(r_{31})\xi_{pk}(r_{12})]$ with $Q_k = Q_h^3$ even for a general threshold function (though some other terms may appear in this case). He suggested that this relation can be used to test the hypothesis that clusters form at high peaks of the primordial density field.

Studies based on the projected catalog of Abell clusters, however, indicate that the three-point correlation function $\zeta_{cl}$ of clusters approximately obeys the hierarchical model with $Q_k = 0$ and $Q_h \approx 0.6$ (Jing & Zhang 1989; Tóth, Hollósi & Szalay 1989). The same hierarchical model seems to hold for $\zeta_{cl}$ of clusters selected from the Lick catalog (Borgani, Jing & Plionis 1992). A few workers have tried to extract the three-point function of clusters from redshift samples of clusters, however the constraints on the form of $\zeta_{cl}$ from these studies are rather weak because the redshift samples are too small (Jing & Valdarnini 1991; Davies & Coles 1993). Recent analyses of the skewness $S_3(R)$ for the Abell clusters and the APM clusters seem to support the hierarchical model because the measured $S_3(R)$ depends very weakly on $R$ (Plionis & Valdarnini 1995; Cappi & Maurogordato 1995; Gaztañaga, Croft & Dalton 1995 hereafter GCD95).

In real observations of $\zeta_{cl}$, as Coles & Davies (1993, hereafter CD93) pointed out, the condition $\xi_m \ll 1$ can be hardly satisfied because the two- and three-point correlation functions of clusters are very difficult to measure on the scales of $\xi_m \ll 1$. They therefore, using numerical integrations, investigated the form of $\zeta_{pk}$ as a function of $\xi_{pk}$ for the quasi-linear regime $\xi_m < 1$. They find that the hierarchical model is a better fit to $\zeta_{pk}$ than the Kirkwood model though neither model provides a good fit to $\zeta_{pk}$. As they pointed out, one problem with this peak theory is that a *region* with density exceeding some threshold may be identified with several distinct clusters because of its finite spatial extent, thus overestimating the correlations on small scales (Coles 1989; Lumsden, Heavens & Peacock 1989). They therefore studied the three-point correlation of local maxima of the density field. It is found that the hierarchical form can approximately describe the three-point correlation function of local maxima. However, because of the technical complexity of this approach, their conclusion is drawn only for the one-dimensional density field. One common prediction of both theories is that the amplitude $Q$ of the three-point function increases with the density threshold.

The main weakness of these analytical theories is that they are unable to treat the non-linear density fluctuations, especially merging of clusters. As pointed out by Croft & Efstathiou (1993) based on N-body simulations, these theories predict too high a two-point correlation function for clusters. This weakness should also influence the calculations of the three-point correlation function. Therefore it is obviously very important to use N-body simulations to calculate the three-point correlation function for clusters. Such a study is not only useful for testing the analytical predictions but also provides a basis to compare observations with theories.

Here we report our results of the three-point correlation function $\zeta$ of clusters in N-body simulations. The simulations are a set of $P^3M$ simulations of $10^6$ particles in a simulation box of 300 or 400 $h^{-1}$Mpc. A description of these simulations will be given in Section 2, where identification procedures of clusters will be discussed. In Section 3, we will report the three-point correlation function of simulated clusters, with emphasis on the dependence of $\zeta$ on sizes and shapes of triangles. We will find that neither the hierarchical nor the Kirkwood model can provide a reasonable fit to the three-point function of simulated clusters. There we suggest another model which can fit $\zeta$ quite well. Then in Section 4, we compare our results with previous works: observational statistics, analytical theories, and simulation studies of the skewness $^\star$. Our main conclusions are summarized in Section 5.

## 2   N-BODY SIMULATIONS

The simulations used here are two sets of large $P^3M$ N-body simulations. The first set (hereafter LCDM) simulates a low-density flat universe of $\Omega_0 = 0.3$, $h = 0.75$ and $\sigma_8 = 1$, where $\Omega_0$ is the current density parameter, $h$ is the Hubble constant in unit of 100 km s$^{-1}$Mpc$^{-1}$ and $\sigma_8$ is the linearly evolved *rms* mass fluctuation in a sphere of $8\,h^{-1}$Mpc at the present time. For this simulation, we assume that the primordial power spectrum is a Harrison-Zel'dovich spectrum. Its linear transfer function is taken from Bardeen et al (1986). The second set (hereafter PIM) simulates an Einstein-de-Sitter universe with $\sigma_8 = 0.8$ and a phenomenological linear power spectrum which was obtained by Branchini, Guzzo & Valdarnini (1994) from an analysis of redshift surveys of galaxies. Because of these parameters, both models provide good approximations to the real universe, although, as Branchini et al. noted, the latter model slightly lacks large-scale clustering power when compared with the APM and EDSGC angular two-point functions of galaxies (Maddox et al. 1990; Collins, Nichol & Lumsden 1992).

The standard $P^3M$ technique is used for these simulations (Hockney & Eastwood 1981; Efstathiou et al. 1985; Valdarnini & Borgani 1991; Jing & Fang 1994). For both models, $10^6$ particles are used in each simulation and 8 realizations are run for each model. For the LCDM simulation, we use a box of $L = 400\,h^{-1}$Mpc and the force resolution $\eta = 0.2\,h^{-1}$Mpc; for the PIM simulation we use $L = 300\,h^{-1}$Mpc and $\eta = 0.3\,h^{-1}$Mpc. Since we have simulated a huge volume of the model universes ( 27 or 64 $\times 10^6\,h^{-3}$Mpc$^3$) with a large number of particles, we are able to calculate accurately the three-point correlation

---

$^\star$ Because the skewness depends on the three-point correlation through an integral, the shape dependence we find in this paper cannot be found in the skewness analysis

function of clusters in these models.

Clusters are selected according to the single criterion — the mass overdensity $\Gamma$ in a spherical volume of radius $r_c = 1.5\,h^{-1}$Mpc. Similar criteria have been used for cluster identification in real observations (Abell 1958; Dalton et al. 1992; Lumsden et al. 1992). In order to study the richness dependence of the cluster three-point correlation function in detail, we have selected both rich and poor clusters with $\Gamma \geq \Gamma_{min}$. $\Gamma_{min}$ is taken to be 90 for the LCDM model and 38 for the PIM model. With these overdensity thresholds, we can identify $\lesssim 3200$ clusters in each realization of the simulation for both models. Moreover each cluster in both models contains at least 20 particles. The poorest clusters in our simulations may be slightly poorer than the clusters in the above mentioned observations. We have used two methods to identify these overdensity regions (i.e. clusters). The first method is based on the *friends-of-friends* algorithm and a description of this method can be found in Jing et al. (1993; see also White et al. 1987). The linking parameter $b$ in this work is taken to be 0.1 times the mean particle separation. This method has been applied only to the LCDM simulation because it is too CPU time-consuming for large simulations ($10^6$ particles here). Therefore we developed another method which can work much faster.

In the second method, we place a grid of $N_g^3$ cells in the simulation volume. $N_g$ is 256 for the LCDM and 192 for the PIM simulations, so that the cell size in both simulations is $1.56\,h^{-1}$Mpc. We count the number $C$ of particles in each cell. If a cell has the count $C$ larger than the count in every of its 26 neighboring cells, this cell is recognized as a local density maximum and the cell center is regarded as the position of the maximum. For each local maximum, we calculate the barycenter $r_0^b$ and the count $C_0$ of the particles inside a sphere of radius $r_c = 1.5\,h^{-1}$Mpc centered at the maximum. Around the barycenter $r_0^b$, we draw a new sphere of $r_c$ and calculate the barycenter $r_1^b$ and the count $C_1$ for the particles in this sphere. We repeat the same calculation for each updated barycenter, until the new count $C_i$ is not larger than the previous one $C_{i-1}$. The previous barycenter $r_{i-1}^b$ and count $C_{i-1}$ are then the center and particle count of a new cluster if its overdensity is larger than the threshold $\Gamma_{min}$. This iterative calculation is attempted to minimize the effect of the uniform grid cells. Some clusters identified in this way may overlap, i.e. the separation between two clusters may be smaller than $2r_c$. We correct this simply by eliminating the smaller one of two overlapping clusters. A sample of clusters has thus been constructed. The cluster sample may still slightly depend on how the grid is placed. Especially some clusters may not appear as local maxima in this particular grid placement, thus they are not included in the sample. In order to reduce this effect, we displace the grid by $1/2$ cell size along the x-axis and/or the y-axis and/or the z-axis (i.e. 7 displacements), and produce, in the way just described, 7 samples of clusters using these displaced grids. Although most clusters ($\sim 90\%$) are common to all the 8 samples, a small fraction of clusters appear only in some of these samples. Therefore we merge the 8 samples, and eliminate the overlapping clusters in the same way as we do for a single sample. This merged sample of non-overlapping clusters is our final sample of clusters.

We have applied the second method both to the LCDM and to the PIM simulations. In order to test whether cluster properties, especially the results of this work, depend on the method of cluster identification, we have compared the mass functions and the two-point correlation functions of the LCDM clusters identified by the two different methods. The two identification procedures produce a consistent result for both measures. In Section 3, we will further show that these procedures give a consistent three-point correlation function of clusters. In the rest of the paper, unless explicitly stated, clusters used for our analysis are those identified by the second method.

## 3 THE THREE-POINT CORRELATION FUNCTION OF CLUSTERS

In order to study the richness dependence of the three-point correlation function, we construct 4 subsamples of clusters for each model according to cluster richness. The $i$-th subsample ($i = 1, 2, 3, 4$) is made of the $400 \times i$ most massive clusters in each realization of the simulation, thus contains a total of $3200 \times i$ clusters. For simplicity, we will call the $i$-th subsample of the LCDM clusters the LCDM$i$ sample and the $i$-th subsample of the PIM clusters the PIM$i$ sample. The overdensity thresholds $\Gamma_{min}$ of the LCDM samples are respectively 90, 136, 195, and 260, and the $\Gamma_{min}$ of the PIM samples are respectively 38, 59, 90, and 125.

### 3.1 The method of estimating the three-point correlation function

The three-point correlation function $\zeta(r_{12}, r_{23}, r_{31})$ is defined by writing the joint probability $dP_{123}$ of finding three clusters concentrated in each of three volume elements $dV_1$, $dV_2$ and $dV_3$ separated by $r_{12}$, $r_{23}$ and $r_{31}$ as,

$$dP_{123} = \bar{n}^3[1 + \xi(r_{12}) + \xi(r_{23}) + \xi(r_{31}) + \zeta(r_{12}, r_{23}, r_{31})]dV_1\,dV_2\,dV_3 \tag{2}$$

where $\bar{n}$ is the mean number density of clusters. Usually it is convenient to use variables $r$, $u$, $v$ instead of $r_{12}$, $r_{23}$, $r_{31}$ (Peebles 1980). For $r_{12} < r_{23} < r_{31}$, the relations of these variables are given as following:

$$r = r_{12}, \qquad u = \frac{r_{23}}{r_{12}}, \qquad v = \frac{r_{31} - r_{23}}{r_{12}}. \tag{3}$$

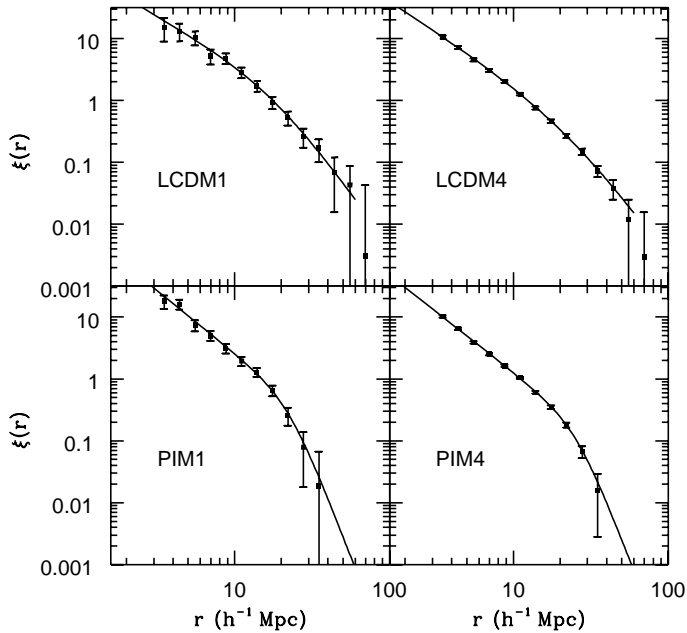

**Figure 1** – The two-point correlation functions of four samples. The solid curves are the best fits of the models described in the text.

The count $N^T(i,j,k)$ of cluster triplets with sides in the $(i,j,k)$-bin $R_i < r \leq R_{i+1}$, $U_j < u \leq U_{j+1}$ and $V_k < v \leq V_{k+1}$ are

$$\begin{aligned}
N^T(i,j,k) &= N^{(1)}(i,j,k) + N^{(2)}(i,j,k) + N^{(3)}(i,j,k) \\
N^{(1)}(i,j,k) &= 8\pi^2 \bar{n}^2 N_{cl} \int_{R_i}^{R_{i+1}} \int_{U_j}^{U_{j+1}} \int_{V_k}^{V_{k+1}} r^5 (u+v) u \, dr \, du \, dv \\
N^{(2)}(i,j,k) &= 8\pi^2 \bar{n}^2 N_{cl} \int_{R_i}^{R_{i+1}} \int_{U_j}^{U_{j+1}} \int_{V_k}^{V_{k+1}} r^5 (u+v) u \left[ \xi(r) + \xi(ru) + \xi(ru+rv) \right] dr \, du \, dv \\
N^{(3)}(i,j,k) &= 8\pi^2 \bar{n}^2 N_{cl} \int_{R_i}^{R_{i+1}} \int_{U_j}^{U_{j+1}} \int_{V_k}^{V_{k+1}} r^5 (u+v) u \, \zeta(r, ru, ru+rv) \, dr \, du \, dv
\end{aligned} \quad (4)$$

where $N_{cl}$ is the total number of clusters in the sample. For the N-body simulations where the periodic boundary is assumed, the above equation is strictly valid for the triplet count if the periodic effect is taken into account (in contrast with real observations where the boundary effect is more difficult to correct). If the two-point correlation function and the triplet count $N^T(i,j,k)$ have been measured, the three-point correlation function $\zeta(r,u,v)$ can be determined through Eq.(4).

We therefore first measure the two-point correlation function $\xi$ for each sample. In Figure 1, we plot, as examples, the two-point correlation function for four samples. The error bars $\sigma_\xi$ shown in the figure are the bootstrap errors estimated by the analytical formula of Mo, Jing & Börner (1992). For the LCDM clusters, we find that the two-point correlation function can be accurately fit by

$$\xi(r) = \frac{1}{ar^{1.5} + br^2 + cr^{3.5}}. \quad (5)$$

Because the PIM model has relatively less power than the LCDM model, the following formula

$$\xi(r) = \frac{1}{ar^{1.5} + br^2 + cr^{6.5}} \quad (6)$$

can nicely fit the two-point correlation function of the PIM clusters. The fitting parameters $a$, $b$, and $c$ are determined for each sample by a least-square fit to $\xi^{-1}(r)$ with the weight $\xi^2 \sigma_\xi^{-1}$. The fitted $\xi$ are shown as solid curves in Figure 1. The figure (and a check with the fit results of other samples not plotted in the figure) shows that these fitting formulae can accurately express the two-point functions of the LCDM clusters for $r < r_{max} \approx 45 \, h^{-1}$Mpc and of the PIM clusters for $r < r_{max} \approx 35 \, h^{-1}$Mpc. These formulae will be used to predict $N^{(2)}(i,j,k)$. We have compared our cluster-cluster $\xi$ of the LCDM model with that of Croft & Efstathiou (1993) of a similar model, and found a good agreement between the two studies. Our results also confirm the weak richness-dependence of the $\xi$ amplitude originally found by Croft & Efstathiou (1993).

We then count the number $N^T(i,j,k)$ of cluster triplets in each sample. Here we use the following bins of $r$, $u$, and $v$:

$$R_i < r \leq R_{i+1} \qquad [\log(R_{i+1}/R_i) = 0.1,\ i = 1,\cdots,i_{max},\ R_1 = 3.1\,h^{-1}\mathrm{Mpc}];$$
$$U_j < u \leq U_{j+1} \qquad [\log(U_{j+1}/U_j) = 1/6,\ j = 1,\cdots,6,\ U_1 = 1]; \qquad (7)$$
$$V_k < v \leq V_{k+1} \qquad [V_{k+1} - V_k = 0.2,\ k = 1,\cdots,5,\ V_1 = 0].$$

Since the two-point correlation function $\xi$ is detected and fit well only at $r \leq r_{max}$, we limit our analysis to triangles with their longest sides smaller than $r_{max}$. Here we take $i_{max} = 12$ ($r_{max} = 39\,h^{-1}\mathrm{Mpc}$) for the LCDM clusters and $i_{max} = 11$ ($r_{max} = 31\,h^{-1}\mathrm{Mpc}$) for the PIM clusters. Because of the constraint $r_{31} \leq r_{max}$, the integral upper limits of $u$ and $v$ for the $N^{(i)}(i,j,k)$ now become $\min(U_{j+1}, r_{max}/r)$ and $\min(V_{k+1}, r_{max}/r - u)$.

We will use the scaled function

$$Q(r,u,v) = \frac{\zeta(r,u,v)}{\xi(r)\xi(ru) + \xi(ru)\xi(ru+rv) + \xi(ru+rv)\xi(r)} \qquad (8)$$

to express the three-point correlation function. For the hierarchical model of $\zeta$, $Q(r,u,v)$ is a constant; for the Kirkwood model ($Q_h \approx Q_k$ in Eq. 1), $Q(r,u,v)$ decreases rapidly with $r$ for $\xi(r) > 1$. Since it is difficult to show the full (3-D) dependence of $Q(r,u,v)$ on $r$, $u$ and $v$ by simple means (e.g. plots), in the following we will separately show the dependence of $Q(r,u,v)$ on each variable by taking an average of $Q(r,u,v)$ over the other two variables. The least-square technique is used to find these averages $\bar{Q}(y)$ ($y = r, u, v$). For example, $\bar{Q}(r)$ is found by minimizing

$$\chi^2 = \sum_{j,k} \left( \frac{N^T(i,j,k) - N^{(1)}(i,j,k) - N^{(2)}(i,j,k) - \bar{Q}(i)N^{(4)}(i,j,k)}{\sigma_{N^T}(i,j,k)} \right)^2 \qquad (9)$$

where $\sigma_{N^T}(i,j,k)$ is the bootstrap error of $N^T(i,j,k)$ estimated by the analytical formula of Mo et al. (1992), and

$$N^{(4)}(i,j,k) = 8\pi^2 \bar{n}^2 N_{cl} \int_{R_i}^{R_{i+1}} \int_{U_j}^{U_{j+1}} \int_{V_k}^{V_{k+1}} r^5(u+v)u[\xi(r)\xi(ru) + \xi(ru)\xi(ru+rv) + \xi(ru+rv)\xi(r)]\,dr\,du\,dv. \qquad (10)$$

This least-square technique will also be used to test the hierarchical model and the Kirkwood model of $\zeta$ and to estimate their parameters $Q$, $Q_h$ and $Q_k$. For the Kirkwood model, we minimize

$$\chi^2 = \sum_{i,j,k} \left( \frac{N^T(i,j,k) - N^{(1)}(i,j,k) - N^{(2)}(i,j,k) - Q_h N^{(4)}(i,j,k) - Q_k N^{(5)}(i,j,k)}{\sigma_{N^T}(i,j,k)} \right)^2 \qquad (11)$$

with

$$N^{(5)}(i,j,k) = 8\pi^2 \bar{n}^2 N_{cl} \int_{R_i}^{R_{i+1}} \int_{U_j}^{U_{j+1}} \int_{V_k}^{V_{k+1}} r^5(u+v)u\xi(r)\xi(ru)\xi(ru+rv)\,dr\,du\,dv. \qquad (12)$$

For the hierarchical model, we just minimize the $\chi^2$ of Eq.(11) with $Q_h$ replaced by $Q$ and with $Q_k$ set to zero.

### 3.2 The results

In Figure 2, we show the results of $\bar{Q}(r)$, $\bar{Q}(u)$ and $\bar{Q}(v)$ of the LCDM clusters. The most noticeable point is the strong dependence of $\bar{Q}(v)$ on $v$. $\bar{Q}(v)$ increases with $v$, and its value is about 0.2 at $v = 0.1$ and about 1.8 at $v = 0.9$. The dependences of $\bar{Q}$ on $r$ and $u$ are much weaker: within $\sim 1\sigma$ error bar, a constant value $\sim 0.5$ is acceptable for $\bar{Q}(r)$ and $\bar{Q}(u)$. The decline of $\bar{Q}(r)$ at large $r$ is due to the facts that $v$ of the triangles with large $r$ should be small because of the longest side $r_{31} < r_{max}$ and that $\bar{Q}(v)$ is an increasing function of $v$. For the range of cluster richness studied in this work, these dependences of $\bar{Q}$ do not depend on the richness, as shown by different rows of the figure. The uncertainty of $\bar{Q}$ of the LCDM1 sample is large because the number of clusters in this richest sample is relatively small.

The results of the PIM clusters are shown in Figure 3. The $\bar{Q}(r)$, $\bar{Q}(u)$ and $\bar{Q}(v)$ of the PIM clusters are nearly the same as those of the LCDM clusters. This might indicate that these $\bar{Q}$-functions do not sensitively depend on the density parameter $\Omega_0$ or the cosmological constant $\Lambda_0$ (these two parameters of the LCDM and the PIM models differ significantly). Since the difference of the linear power spectra between the two models is not large, it is difficult to say whether these functions depend on the power spectrum or not.

We have also done the same analysis for the LCDM clusters identified by the *friends-of-friends* algorithm (Method I). The results are shown in Figure 4. The three $\bar{Q}$-functions of this figure are nearly the same as those in Fig. 2. This means that our results are robust, not depending on the method used to identify clusters.

The weak $r$-dependence and the strong $v$-dependence of $\bar{Q}$ means that neither the Kirkwood model (with $Q_h = Q_k$) nor the hierarchical model is a good model for the three-point correlation function of clusters (for the range of scales analyzed here). We propose another empirical model in which $Q(r,u,v)$ depends only on $v$ through the relation

$$Q(r,u,v) = \Theta 10^{1.3v^2}. \qquad (13)$$

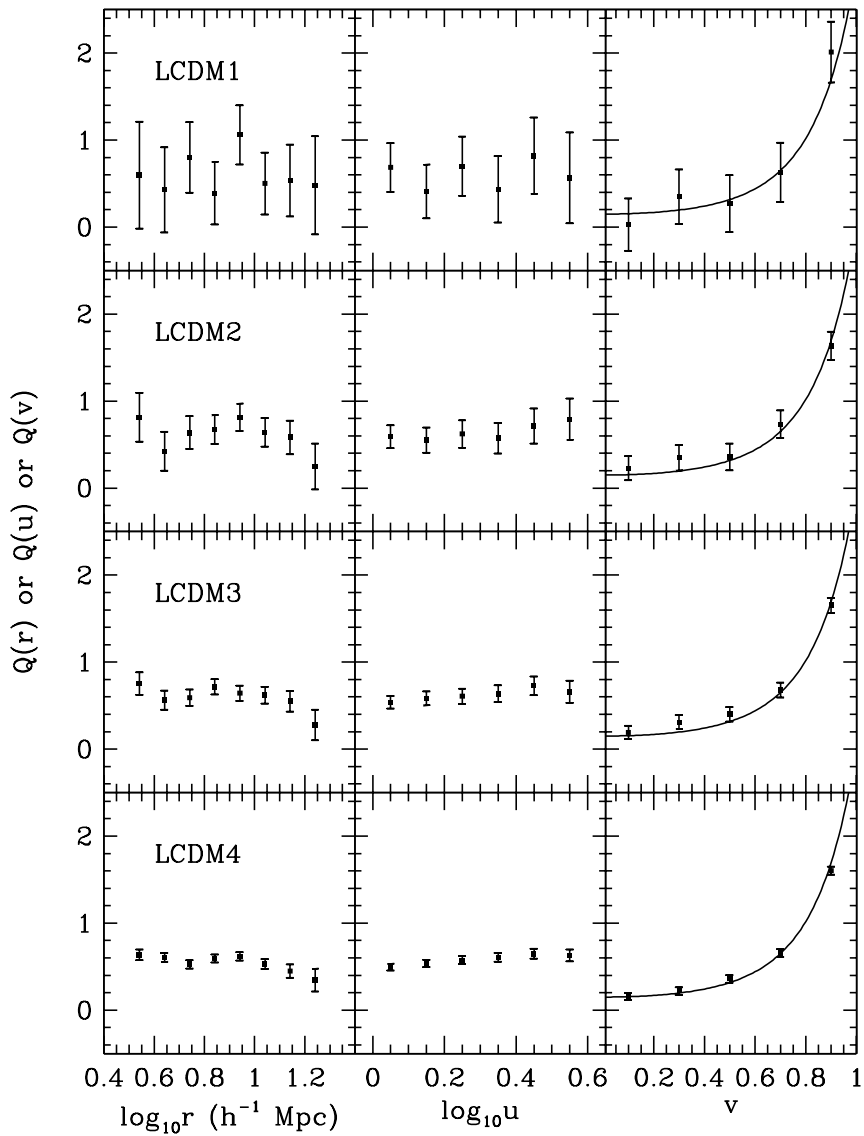

**Figure 2** – The averaged $\bar{Q}$ of the LCDM clusters as a function of $r$ (the left panels), $u$ (the middle panels) or $v$ (the right panels). Each row of three panels shows the results of one sample, with the sample name given in the right panel. The solid curves in the left panels are $\bar{Q}(v) = 0.15 \, 10^{1.3v^2}$.

As shown by the $\chi^2$ tests, this model (hereafter Model III) can fit the simulation data of $\zeta$ much better than the Kirkwood and the hierarchical models. Table 1 lists $\chi^2_{min}$ of these three model fits to $\zeta$ of the LCDM clusters (cf. Eq. 11). The $\chi^2_{min}$ of Model III is much smaller than those of the other two models, which means that Model III can better fit the data of $\zeta$ than the other two models. For 239 to 240 degrees of freedom, the $\chi^2_{min}$ of the LCDM4 model is 71 for Model III, 561 for the hierarchical model, 671 for the Kirkwood model with $Q_h = Q_k$, and 544 for the Kirkwood model with two free parameters $Q_h$ and $Q_k$, therefore the hierarchical and the Kirkwood models fail to describe the simulation data. Since the absolute values of $\chi^2_{min}$ depend on the error model (here the bootstrap error) used in the least-square fits, we caution readers against over-interpreting the statistical meaning of these absolute values, though the bootstrap error was shown to be a good error model for such fits (see Mo et al. 1992 for a very detailed discussion). The best values of $\Theta$ from these fits are given in Table 2. $\Theta$ is about 0.15 for the LCDM clusters and 0.13 for the PIM clusters. The solid curves in Figs. 2 – 4 are the Model III predictions with these $\Theta$ values. It is clear that Model III can very nicely fit the simulation data of $\zeta$.

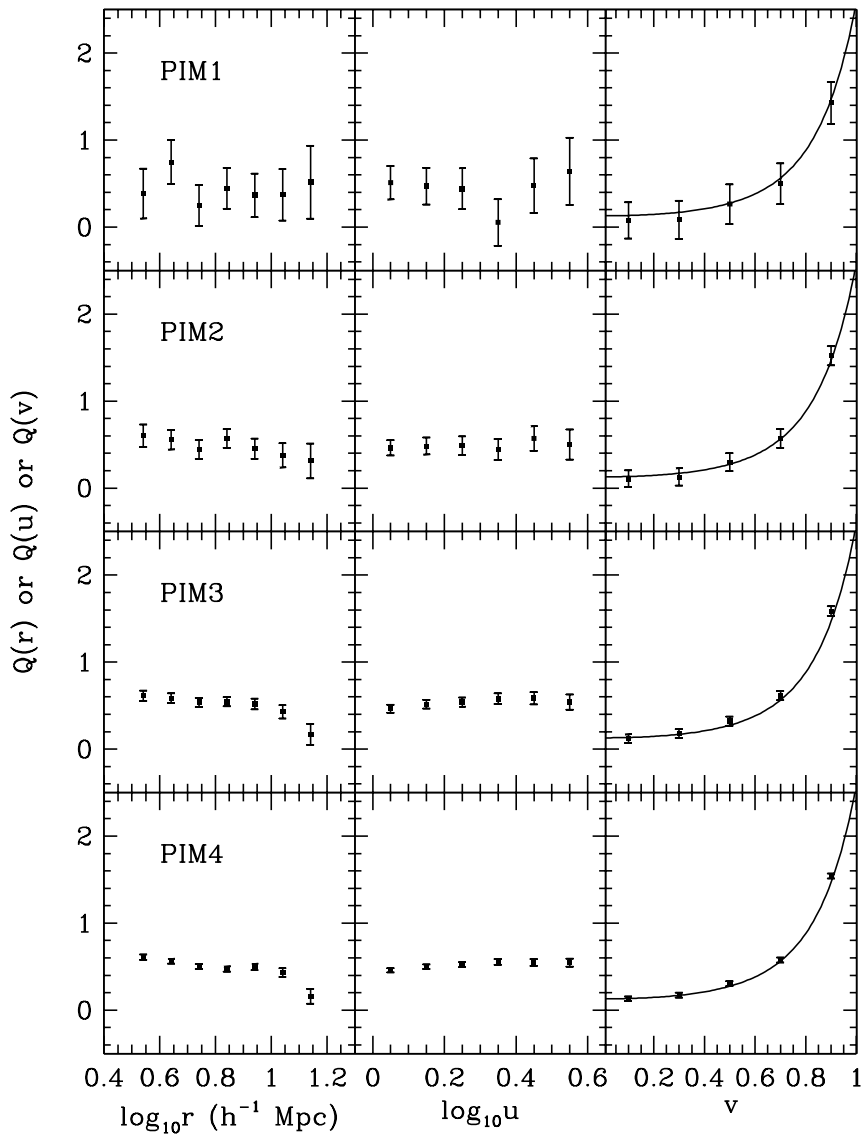

**Figure 3** – The averaged $\bar{Q}$ of the PIM clusters as a function of $r$ (the left panels), $u$ (the middle panels) or $v$ (the right panels). Each row of three panels shows the results of one sample, with the sample name given in the right panel. The solid curves in the left panels are $\bar{Q}(v) = 0.13\ 10^{1.3v^2}$.

## 4 DISCUSSION

First let us compare our N-body results with the predictions of some analytical theories of cluster formation (CD93). The analytical theories explore the clustering properties either of regions with density fluctuation above some threshold (hereafter the peak theory) or of local linear density maxima above some threshold (hereafter the maxima theory). The peak theory predicts that $Q$ slightly decreases with the increase of $v$ (cf. Figs. 1–2 of CD93), contrary to our N-body results. The maxima theory can predict the three-point correlation function only for the one-dimensional case (i.e. $v = 1$), so it is unknown how $Q$ changes with $v$ in this theory. Both theories predict that $Q$ increases with the peak (maxima) threshold, which is inconsistent with our result that $Q$ does not depend on the richness of clusters. The failure of the analytical theories to match our N-body results implies that these theories can not be used to predict the three-point correlation function for clusters. It is not very surprising that the analytical theories are inconsistent with our N-body results, since these theories have also met problems to predict the two-point correlation function of clusters (Croft & Efstathiou 1993).

We have noted two recent works which have analyzed the three-point correlation function and/or skewness $S_3(R)$ for rich clusters in N-body simulations. Watanabe, Matsubara, & Suto (1994) analyzed the three-point correlation function and $S_3(R)$ for simulation clusters. However, their cluster samples are too small, so that they are unable to address the dependences of $Q$

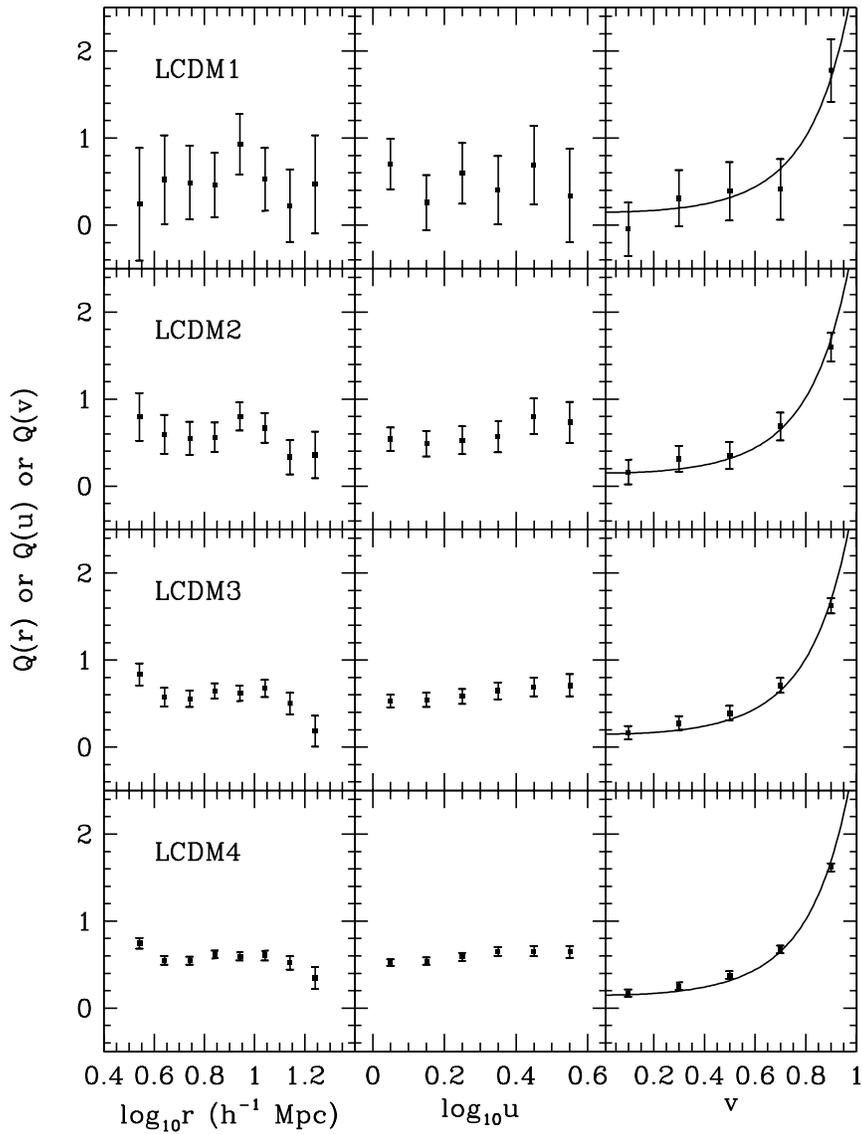

**Figure 4** – The same as Fig. 2, but the clusters are identified by the *friends-of-friends* algorithm.

**Table 1** $\chi^2_{min}$ of three model fits to the LCDM samples

|  | Hierarchical | Kirkwood 1 | Kirkwood 2 | Model III |
| --- | --- | --- | --- | --- |
| LCDM1 | 51.0 | 54.4 | 50.4 | 37.5 |
| LCDM2 | 80.0 | 94.1 | 75.4 | 43.4 |
| LCDM3 | 182.2 | 223.5 | 178.4 | 52.3 |
| LCDM4 | 560.8 | 671.1 | 543.8 | 71.4 |

The number of data points for these fits are 241. In Kirkwood 1, $Q_h = Q_k$ is assumed. In Kirkwood 2, both $Q_h$ and $Q_k$ are free fit parameters.

**Table 2** The fit values of $\mathcal{Q}$ for the LCDM and PIM clusters

| Sample No. | LCDM | PIM |
| --- | --- | --- |
| 1 | $0.16 \pm 0.03$ | $0.11 \pm 0.02$ |
| 2 | $0.15 \pm 0.02$ | $0.13 \pm 0.01$ |
| 3 | $0.15 \pm 0.01$ | $0.136 \pm 0.005$ |
| 4 | $0.140 \pm 0.004$ | $0.131 \pm 0.003$ |

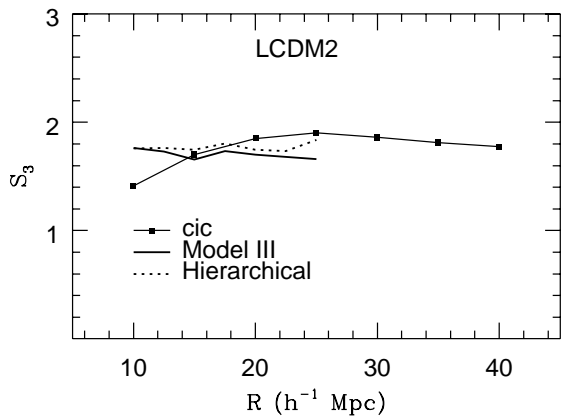

**Figure 5** – The skewness $S_3(R)$ of an LCDM cluster sample measured with the count-in-cell (CIC) analysis (squares connected with solid lines), compared with the integral results of Equation (14) for the Hierarchical model (dotted line) and for Model III (thick solid line). The value at $R = 10\,h^{-1}$ Mpc given by the CIC analysis is slightly low due to the finite size of the clusters. The numerical integration accuracy of $S_3(R)$ is controlled to $\sim 10\%$. Within this accuracy, the integral results given by both models are in good agreement with the result from the CIC analysis.

on $r$, $u$ and $v$, though their average value of $Q$ is consistent with our results.

GCD95 have analyzed only the skewness of clusters (but not the three-point function) in a set of CDM simulations. The normalized skewness $S_3(R)$ of a spherical volume is related to the three-point correlation function through

$$
\begin{aligned}
S_3(R) &= \frac{\bar{\zeta}(R)}{\bar{\xi}^2(R)} \\
\bar{\zeta}(R) &= \frac{1}{V^3} \int_{\text{sphere}\,R} d\bm{r}_1\,d\bm{r}_2\,d\bm{r}_3 \zeta(r_{12}, r_{23}, r_{31})\,, \\
\bar{\xi}(R) &= \frac{1}{V^2} \int_{\text{sphere}\,R} d\bm{r}_1\,d\bm{r}_2\, \xi(r_{12})
\end{aligned}
\qquad (14)
$$

where $V = 4\pi R^3/3$. We have measured $S_3(R)$ for our LCDM cluster samples by the count-in-cell analysis (Peebles 1980) with a result in good agreement with that of an LCDM model of GCD94. Our measured $S_3(R)$ is also nicely consistent, as expected, with the $S_3(R)$ numerically calculated through Eq.(14) using Model III. This is shown in Figure 5. $S_3(R)$ is almost a constant $\sim 1.7$ for $R \leq 25\,h^{-1}$ Mpc (we limited our analysis to $R \leq 25\,h^{-1}$ Mpc because the three-point correlation function is measured only for $r_{31} \leq 40\,h^{-1}$ Mpc. The constant $S_3(R)$ might extend much beyond $R = 25\,h^{-1}$ Mpc.).

A constant $S_3(R)$ was usually regarded as a support for the hierarchical model in the literature. Actually the hierarchical model with $Q \approx 0.55$ can reproduce, using Eq.(14), our measured $S_3(R)$ (see Fig. 5), but the hierarchical model clearly fails to account for the direct count $N^T(i,j,k)$. It is not surprising that the different models for $\zeta$ can lead to the same result for $S_3(R)$, since $S_3(R)$ is an integral of $\zeta$. Clearly a constant $S_3(R)$ is not sufficient to argue for the hierarchical model. The count-in-cell analysis (or the moment method) has been widely used to study high-order correlation functions of extragalactic objects. Although these analyses have already yielded useful information for high-order correlations (e.g. Plionis & Valdarnini 1995; Cappi & Maurogordato 1995; GCD95), the strong $v$-dependence of $\bar{Q}$ found in this work will be definitely lost in the moment method.

The average value of $\bar{Q}$ (i.e. averaged over $r$, $u$ and $v$) of our model clusters is about 0.55, consistent with the observational results from the statistical analyses of Abell clusters, Lick clusters and APM clusters (Jing & Zhang 1989; Jing & Valdarnini 1991; Borgani et al. 1992, Tóth et al. 1989; Plionis & Valdarnini 1995; Cappi & Maurogordato 1995; GCD95). Of these observational works, only JZ89 have examined the richness-dependence and the $v$-dependence of $Q$. They analyzed the three-point function for Abell clusters of richness $R \geq 1$ and $R \geq 2$ respectively, and found that $Q$ does not depend on the richness of clusters. Their result is consistent with our results here. As for the $v$-dependence, because the Abell catalogue is a projected sample, they measured the value of $q$ in two bins of $0 < v < 0.5$ and $0.5 < v < 1$ [where $q(\theta, u, v)$ is defined in the same way as $Q$ but with the angular two and three-point correlation functions, see JZ89 for details]. They found that $q$ does not strongly depend on $v$, at least much more weakly than the $v$-dependence of $\bar{Q}$ found here. The projection effect in a two-dimensional catalogue may have reduced the $v$-dependence. We have examined this problem using the Limber equation (Peebles 1980). Assuming that the three-point correlation function obeys Model III with $\Theta = 0.15$, the two-point function $\xi \propto r^{-2}$ and the cluster sample is a perfect distance-limited sample of redshift $z = 0.2$, we calculate $q(\theta, u, v)$ for different values of $\theta$, $u$ and $v$. We find that $q$ depends very weakly on $\theta$ and $u$, but does depend on $v$. The $v$-dependence of $q$ is shown in Figure 6. We see that $q$ increases from $q \approx 0.5$ at $v = 0.1$ to $q \approx 0.9$ at $v = 0.9$. The $v$-dependence of $q$ in the two-dimensional case is much weaker than that of $Q$ in the three-dimensional case. From Figure 6, we expect that the value of $q$ in the two bins of $0 < v < 0.5$ and $0.5 < v < 1$ is about 0.5 and 0.7 respectively. These two values are consistent with the result of JZ89 within the statistical uncertainty of their work ($\Delta q \approx 0.2$). So it is not surprising that they could not discover the $v$-dependence of $q$ in their work. Overall the observational results are not contradictory to our result based on N-body simulations.

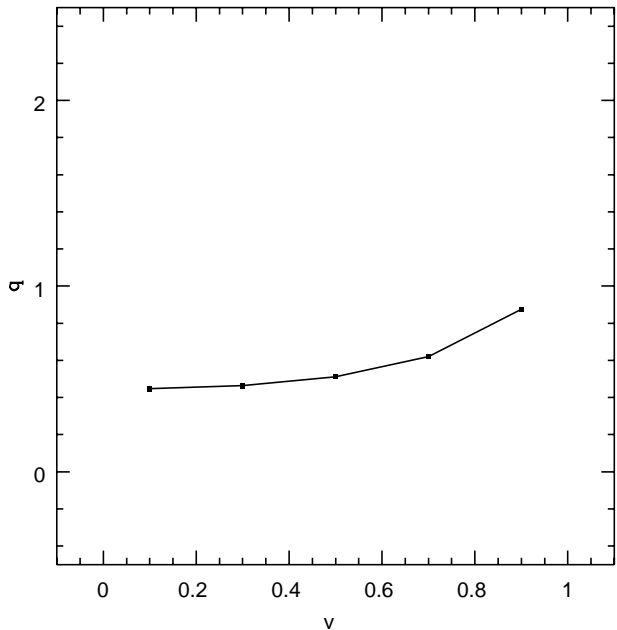

**Figure 6** – The $v$-dependence of $q$ predicted by the Limber equation for a projected catalogue of rich clusters, if the cluster three-point correlation obeys the model of this paper.

It is worth emphasizing that the analytical theories compared above are simple theories of cluster formation. Their failure to explain our N-body results only means the limitation of these simple theories. A similar $v$-dependence to that found here has been predicted for the *mass* three-point function $\zeta_m$ of the initially Gaussian fluctuation field by the second-order perturbation theory (Fry 1984). Although the relations between the cluster $\zeta$ and the mass $\zeta_m$ are yet unknown, it seems that the non-Gaussian underlying density field (which is resulted from the dynamical evolution of a Gaussian density field) has an important impact on the form of the three-point function of clusters (see Fry & Gaztañaga 1993 for a general discussion of the biasing issue in the second-order perturbation theory). This could also be the reason why the peak theories or the local maxima theories fail to explain our N-body results, since these theories assume that the underlying density field is Gaussian. The relations between $\zeta$ and $\zeta_m$ will be explored in a subsequent paper (Jing et al. in preparation). At this moment, it is reasonable to argue that the $v$-dependence of $Q$ probably is an important feature of clusters formed in gravitational instability theories of Gaussian fluctuation fields. As an example of non-Gaussian models, the toy model of a bubble universe in which galaxies are distributed on shells and clusters form at the "knots" of three crossing shells predict a very weak $v$-dependence of $Q$ for rich clusters (Jing 1991). This toy model is inspired by a non-gravitational instability theory of galaxy formation — the explosion scenario (Ostriker & Cowie 1981). Therefore, it is worthwhile to conduct a statistical analysis for the $v$-dependence of $Q$ in redshift samples of rich clusters. The $v$-dependence of $Q$ could serve as a discriminator between different scenarios of cluster formation.

## 5 CONCLUSIONS

In this paper we have used two sets of N-body simulations to study the three-point correlation function of clusters in theoretical models. We found:

(1) The scaled three-point function $Q(r, u, v)$ depends weakly on $r$ and $v$, but very strongly on $v$. This function is found to be universal for the LCDM clusters and for the PIM clusters. Furthermore, this function does not depend on the cluster richness.

(2) Two well-known models of $\zeta$, the Kirkwood and the Hierarchical model, cannot be accommodated to these features of $Q(r,u,v)$. We proposed another model $Q(r,u,v) = \Theta\, 10^{1.3v^2}$ which can fit the data of $\zeta$ very nicely with $\Theta \approx 0.14$.

(3) Simple analytical theories of cluster formation, like the peak theories or the local maxima theories, fail to explain the weak dependence of $Q(r,u,v)$ on the cluster richness and/or the $v$-dependence of $Q(r,u,v)$ found in our N-body simulations. The reason might be that these theories cannot describe the non-linear density fluctuation, especially the merging processes of clusters and that these theories assume a Gaussian fluctuation for the underlying density field.

(4) The $v$-dependence of $q$ in a projected catalogue of clusters is found to be much weaker than the $v$-dependence of $Q$. This is probably the reason why JZ89 have not found the $v$-dependence in their analysis of $\zeta$ for the Abell catalogue. The weak richness dependence of $Q$ they found, agrees with our N-body results. Overall, our N-body results of $\zeta$ are not contradictory with all observational results currently obtained.

(5) These specific but robust features of the cluster three-point correlation function found in this paper, might be an important feature of Gaussian gravitational instability theories. Simple toy models of the explosion theory of galaxy formation predict a $Q$ which depends rather weakly on $v$. Therefore a statistical study of $\zeta$ for redshift samples of rich clusters would be very important for testing theories of galaxy formation.

## ACKNOWLEDGMENTS

We would like to thank S. Borgani, Z.G. Deng, G. Efstathiou, B. Jain, S. Matarrese, H.J. Mo, L. Moscardini, M. Plionis, and S.D.M. White for helpful conversations, and E. Gaztañaga, the referee of this paper, for helpful comments. YPJ acknowledges the receipt of an Alexander-von-Humboldt research fellowship.